\documentclass[article,aps,prl,twocolumn,groupedaddress,showpacs,showkeys,superscriptaddress,10pt]{revtex4-1}
\usepackage[utf8]{inputenc}
\usepackage{graphicx}
\usepackage[colorlinks]{hyperref}
\usepackage{amssymb}
\usepackage{ctable}
\usepackage{amsmath}
\usepackage{amsfonts}
\usepackage{amssymb}
\begin{document}

\title{Steady motional entanglement between two distant levitated nanoparticles}
\author{Guoyao Li}
	\affiliation{Center for Quantum Technology Research and Key Laboratory of Advanced Optoelectronic Quantum Architecture and Measurements (MOE), School of Physics, Beijing Institute of Technology, Beijing 100081, China}

	\author{Zhang-qi Yin}\email{zqyin@bit.edu.cn}
		\affiliation{Center for Quantum Technology Research and Key Laboratory of Advanced Optoelectronic Quantum Architecture and Measurements (MOE), School of Physics, Beijing Institute of Technology, Beijing 100081, China}
		\affiliation{Beijing Academy of Quantum Information Sciences, Beijing 100193, China}
		
\date{\today}

\begin{abstract}
Quantum entanglement in macroscopic systems is not only essential for practical quantum information processing, but also valuable for the study of the boundary between quantum and classical world. However, it is very challenge to achieve the steady remote entanglement between distant macroscopic systems. 
We consider two distant nanoparticles, both of which are optically trapped in two cavities. Based on the coherent scattering mechanism, we find that the ultrastrong optomechanical coupling between the cavity modes and the 
motion of the levitated nanoparticles could achieve. The large and steady entanglement between the filtered output cavity modes and the motion
of nanosparticles can be generated, if the trapping laser is under the red sideband. Then through entanglement swapping, the steady motional entanglement between the distant nanoparticles can be realized. We numerically simulate  and find that the two nanoparticles with 10 km distance can be entangled for the experimentally feasible parameters, even in room temperature environment.   
\end{abstract}
\maketitle
$\emph{Introduction}.-$ Quantum entanglement is widely considered to be a foundamental resource in quantum computation~\cite{Preskill2018quantumcomputingin}, quantum network~\cite{RevModPhys.82.1209}, and quantum metrology~\cite{degen2017quantum}. The entanglement has been generated in various systems, such as atoms~\cite{hofmann2012heralded,ritter2012elementary}, atomic assembles~\cite{matsukevich2006entanglement,li2021heralding}, nitrogen-vacancy centers~\cite{bernien2013heralded}, superconducting qubits~\cite{2021Natur.590..571Z}, 
and mechanical oscillators~\cite{jost2009entangled,ockeloen2018stabilized,humphreys2018deterministic,riedinger2018remote}.
The deterministic quantum entanglement has been realized between two systems with distance around 1 meter or less~\cite{humphreys2018deterministic,pompili2021realization,ockeloen2018stabilized,palomaki2013entangling}. 
In order to achieve the remote entanglement between distant quantum systems, the pre- and/or post-selections are usually applied with the cost of low success possibility~\cite{hensen2015loophole,li2021heralding,riedinger2018remote,thomas2021entanglement}.  As the decoherence rate is proportional to the size of the system, it is extremely challenging to achieve the steady entanglement between the distant macroscopic systems.

Because of ultra-high Q factor ($>10^{10}$), the levitated optomechanical system is one of the best testbed for macroscopic quantum mechanics~\cite{PNAS2010,NJPh2010,Yin2013,millen2020optomechanics,gonzalez2021levitodynamics}. Many quantum phenomena have been predicted in this macroscopic system, such as quantum superpositions and matter-wave interference~\cite{Isart2011,Yin2013a,Bose2013,Chen2018}, gravity induced entanglement~\cite{Bose2017,Marletto2017}, quantum time crystal~\cite{PhysRevA.102.023113,Huang2018}, etc. 
Recently, the center of mass (CoM) motion the optically levitated nanoparticle has been cooled to quantum ground state~\cite{delic2020cooling,2021Natur.595..378T,2021Natur.595..373M}, which is the first step towards the macroscopic quantum phenomena. 
The strong coupling between the CoM motion of the levitated nanoparticle and the cavity mode has also been achieved via the coherent scattering mechanism~\cite{de2021strong}. 
Inspired by these breakthroughs,
the mechanical squeezing and entanglement for the optically levitated nanoparticle within a single cavity have been theoretically studied~\cite{PhysRevResearch.2.013052,PhysRevA.101.011804,chauhan2020stationary}.

In this letter, we propose a practical scheme to realize steady entanglement between the motional modes of two distant optically levitated nanoparticles, which couple with two cavities \cite{pirandola2006macroscopic,abdi2012entanglement,borkje2011proposal}.
Based on the coherent scattering mechanism ~\cite{gonzalez2019theory}, 
we calculate the optomechanical coupling $g_{s\phi}$ ($g_{sy}$) between the torsional (CoM) modes and the cavity modes. We find that the {\em ultra-strong coupling regime} ($g_{s\phi(sy)}/\omega_{\phi(y)}>0.1$) could achieve for the current experimental conditions ~\cite{2019NatRP}. 
The rotating-wave approximation is no longer valid here. 
If the trapping lasers are in red sideband of the cavity modes, the torsional (CoM) motion of the levitated nanoparticles is not only cooled down, but also strongly entangled with the cavity modes \cite{PhysRevApplied.14.054052}. 
We further propose to unconditionally entangle the torsional (CoM) motion of the distant nanoparticles via entanglement  swapping~\cite{furusawa1998unconditional,spedalieri2013covariance,pirandola2006macroscopic,abdi2012entanglement,borkje2011proposal}.
The resulting steady entanglement between two remote nanoparticles is robust to both thermal noise and photon loss. We numerically calculate the steady entanglement between two levitated nanoparticles with the experimentally feasible parameters~\cite{hoang2016torsional,bang2020five,ahn2018optically}, and find that maximal distance between the nanoparticles could be larger than $10$ km.

\begin{figure}
  \centering
  \includegraphics[height=6cm]{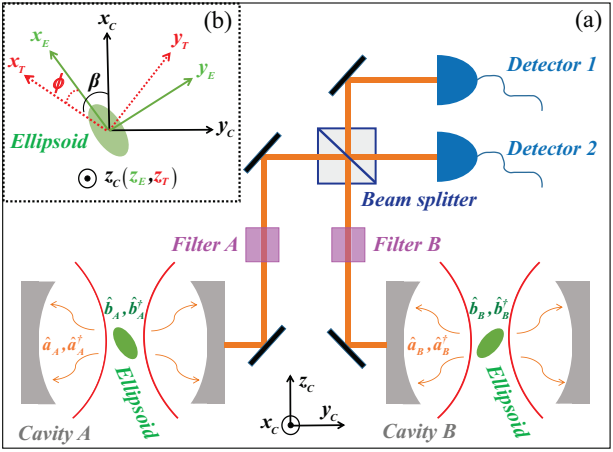}

  \caption{(Color online) (a) Schematic diagram for the scheme that contains two remotely levitated nanoparticles locating in the center of cavity modes. The optomechanical coupling between the cavity mode $\hat a_\xi$ and torsional (or CoM) mode $\hat b_\xi$ is induced through the coherent scattering mechanism. The output modes of $\xi$-th $(\xi=A,B)$ cavity are selected by the filtering operation and then measured by the Bell-like detection. (b) $\{x_E,y_E,z_E\}$, $\{x_C,y_C,z_C\}$, and $\{x_T,y_T,z_T\}$,
 are the Cartesian coordinate systems for the ellipsoid, the cavity mode, and the optical tweezers. The polarization directions of the optical tweezers (cavity mode) align to the long axis of the ellipsoid $x_T$ and $x_C$ respectively. 
  $\phi$ $(\beta)$ is the angel between the $x_T$ ($x_E$) and $x_C$. Optical tweezers propagate along the $z_T$ axis.}
  \label{Fig:1}
\end{figure}

$\emph{Model}.-$
As shown in Fig~\ref{Fig:1}~(a), we theoretically consider two optically levitated nano-ellipsoids locating in distant cavity $\xi$ ($\xi=A,B$).
Each nano-ellipsoid is trapped via linearly polarized optical tweezers.
The dipole moment of the nano-ellipsoid in the optical tweezers is given as
\begin{eqnarray}
\hat{P}=\hat{\alpha}'(\theta,\phi)\hat{E}(r),
\end{eqnarray}
where $\hat{\alpha}'(\theta,\phi)$ is the polarization tensor and $\hat{E}(r)$ denotes the amplitude of the electrical field in position $r$. 
Noted that the motion of the nano-ellipsoid has five degrees of freedom $\{x,y,z,\theta,\phi\}$~\cite{SM}, while the spinning around the long axis of the ellipsoid is neglected due to the symmetry~\cite{bang2020five}.
The parameters 
$\{x,y,z\}$ and $\{\theta,\phi\}$ are corresponding to the CoM position and the orientation position of the ellipsoid, respectively.
In order to minimize the energy of the optical tweezers, the levitated nano-ellipsoids must have both equilibrium position and equilibrium orientation, leading to both CoM modes and torsional modes, respectively~\cite{hoang2016torsional}. 
When the optical tweezers are placed in the node (anti-node) of the cavity field, the coherent scattering coupling between the CoM (torsional) mode and cavity mode is maximized. Under that condition, the CoM and torsional motion can be decoupled from each other (cf., Sec.~III in~\cite{SM}). 
Beside,
due to the lack of the cavity photons, the cavity enhanced coupling between the torsional (CoM) mode and the cavity mode can be safely neglected~\cite{delic2020cooling,SM}.
Therefore, we can independently consider the coherent scattering coupling between the cavity mode and the torsional mode, or CoM mode.

\begin{figure}
  \centering
  \includegraphics[height=6cm]{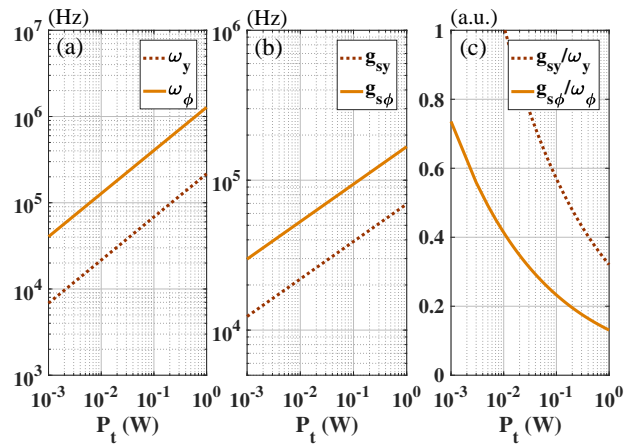}
  \caption{(Color online) The torsional (CoM) frequency $\omega_{\phi(y)}$ (a), the coherent scattering coupling strength between the cavity mode and torsional (CoM) mode $g_{s\phi(sy)}$ (b), and the ratio $g_{s\phi(sy)}/\omega_{\phi(y)}$ (c) as a function of the optical tweezers power $P_t$.
 $\omega_{\phi}$ ($\omega_{y}$) and $g_{s\phi}$ ($g_{sy}$) are theoretically calculated via cavity length $L=1$ mm ($L=10$ mm) and semiaxis of the ellipsoid $a=100$ nm, $b=c=50$ nm ($a=150$ nm, $b=c=60$ nm).
 Other parameters used in this figure are listed as follow: wavelength of the optical tweezers $\lambda_t=1550$ nm, the waist of the tweezers $w_0=1$ $\mu$m, the relative permittivity of the ellipsoid $\varepsilon=2.1$, the density of the ellipsoid $\rho=2200$ kg/m$^3$~\cite{hoang2016torsional,ahn2018optically}.}
  \label{Fig:2}
\end{figure}

The CoM (torsional) mode frequencie $\omega_{\phi(y)}^\xi$ and the optomechanical coupling $g_{s\phi(sy)}^\xi$ are calculated and shown in  Fig.~\ref{Fig:2}. We find that the ratio  $g_{s\phi(sy)}^\xi/\omega_{\phi(y)}^\xi$ is in proportional to  $P_t^{1/4}$, and could be larger than $0.1$ for optical tweezers power in focus $P_t$ from $10^{-3}$ to $1$ W. Therefore,  the optomechanical coupling in our scheme is in the ultrastrong coupling regime~\cite{2019NatRP}. 
For example, if we set $P_t=0.01$ W ($P_t=0.41$ W), the ratio between the coupling strength $g_{s\phi}=53$ kHz ($g_{sy}=56$ kHz) and the motional frequency  $\omega_\phi=128$ kHz $(\omega_y=139$ kHz) is around $0.4$.
The rotating-wave approximation is no longer valid here. 
Both of the rotating and counter-rotating terms of the Hamiltonian must be considered.
The optomechanical Hamiltonian for either CoM mode or torsional mode has the same form ~\cite{SM}
\begin{equation}
\begin{split}
\hat H = \sum\limits_{\xi  = A,B} \big[ {\hbar {\Delta ^\xi }\hat a_\xi ^\dag {{\hat a}_\xi } + \hbar \omega _m^\xi \hat b_\xi ^\dag {{\hat b}_\xi } - \hbar g_{}^\xi ( {\hat a_\xi ^\dag  + {{\hat a}_\xi }} ) ( {\hat b_\xi ^\dag  + {{\hat b}_\xi }} )} \big]
\end{split}
\label{Eq:2}
\end{equation}
where $\omega_{m}^{\xi}$ and $g^\xi$ are the torsional (CoM) frequency and coherent scattering coupling strength between the cavity mode and the torsional (CoM) mode, $\Delta^\xi=\omega_{c}^\xi-\omega_{0}^\xi$ is the detuning between the cavity mode frequency $\omega_c^\xi$ and the optical tweezers frequency  $\omega_0^\xi$.
${\hat a^{\dag}_{\xi} }$ ($\hat b^{\dag}_{\xi}$) and $\hat a_{\xi}(\hat b_\xi)$ are creation and annihilation operators for the cavity mode (torsional or CoM mode), following the commutation relation $\left[ {\hat a_{\xi},{\hat a^{\dag}_\xi }} \right] = 1$ $(\left[ {\hat b^{\dag}_\xi,{\hat b_{\xi} }} \right] = 1)$.

In the ultrastrong coupling regime,
no matter the detuning is on red sideband $\Delta^\xi =\omega^\xi_{m}$ or on blue sideband $\Delta^\xi =-\omega_m^\xi$,  both the beam-splitter (BS) interaction terms $\hat a^{\dag}_\xi\hat b_\xi + \hat a_\xi \hat b^{\dag}_{\xi}$ and the two-mode squeezing (TMS) terms $\hat a_\xi\hat b_\xi + \hat a^{\dag}_\xi \hat b^{\dag}_{\xi}$ in Eq.~\eqref{Eq:2} have to be considered in this letter. 
The motional modes could entangle with the cavity output modes under either red or blue sideband. 
However, on the red sideband $\Delta^\xi= \omega_{m}^\xi$, the motion of levitated nano-ellipsoids can be cooled down and the system becomes much more stable than the blue sideband case. Therefore, in our scheme we adopt the red sideband detuning.

Based on Hamiltonian~\eqref{Eq:2}, the Langevin equations can be written down as follows~\cite{SM}
\begin{equation}
\dot {\hat{a}}_\xi =  - i\Delta^\xi \hat{a}_\xi - \frac{\kappa^\xi }{2}\hat{a}_\xi  + ig^\xi \left( \hat b_\xi + {\hat b_{\xi}^{\dag }} \right) + \sqrt {\kappa^\xi} {\hat a_\xi^{in}},
\label{Eq:3}
\end{equation}
\begin{equation}
\begin{split}
\dot {\hat{b}}_\xi =  - i{\omega _{m}^\xi }\hat{b}_\xi - \frac{\gamma^\xi }{2}\hat{b}_\xi
+ig^\xi \left( {{\hat a^{\dag }_\xi} + \hat{a}_\xi} \right) + 
\sqrt {\gamma^\xi}  \hat b^{{in}}_{\xi},
\end{split}
\label{Eq:4}
\end{equation}
where $\kappa^\xi$ ($\gamma^\xi$) is the decay rate of the mode $\hat{a}_\xi$ ($\hat{b}_\xi$).
The input noise terms are $\hat{a}^\xi_{in}$ and $\hat{b}^{{in}}_{\xi}$, which have the following correlation relations
$\left\langle {\hat a_\xi ^{in}\left( t \right)\hat a_\xi ^{in\dag }\left( {t'} \right)} \right\rangle  = \delta \left( {t - t'} \right)$
and
$\left\langle {{\hat{b}^{{in}}}_{\xi}\left( t \right){\hat{b}_{\xi}^{in{\dag}}}\left( {t'} \right)} \right\rangle  = \left( {\bar n^\xi + 1} \right)\delta \left( {t - t'} \right)$,
respectively~\cite{vitali2007optomechanical}. We denote 
$\bar n^\xi = [\exp(\hbar \omega_{m}^{\xi}/k_B T_{\xi})-1]^{-1}$
as the mean thermal excitation number for the motional mode $\hat{b}_\xi$ at temperature $T_{\xi}$, where $k_B$ is  Boltzmann constant.

From the Eqs.~\eqref{Eq:3} and \eqref{Eq:4}, we can calculate the covariance matrix (CM) $V^{out}$ of the output mode to character the entanglement between the output cavity mode and the motional mode  (cf., Secs.~IV and V in~\cite{SM}).
In order to enhance this entanglement, the bad cavity condition $\kappa^\xi \gg g^\xi$ is also adopted. 
The dynamics of the intracavity field can be adiabatically eliminated, leading to the effectively directly coupling between the motional  mode and output field ~\cite{gut2020stationary}.
However, the output field has broad bandwidth which covers both blue and red sidebands regime.
In order to distinguish those output photons, a filtering operation is introduced to decompose the continuous traveling light into different temporal modes and code it in a time sequence~\cite{gut2020stationary}.
The output photons in different frequency are filtered into either the TMS or BS temporal modes in the Fourier domain~\cite{borkje2011proposal}. In the bad cavity limit,  there is almost no overlap between them. Therefore, the temporal modes for the TMS and BS can be easily distinguished in time~\cite{gut2020stationary}.

Here we define two filtering operations
${F_t^\xi} = \sqrt {2\Gamma^\xi } {e^{\Gamma^\xi t}}{e^{ - i{\omega _{m}^{\xi}}t}}(t\leq0)$ and
${F_b^\xi} = \sqrt {2\Gamma^\xi } {e^{ - \Gamma^\xi t}}{e^{i{\omega _{m}^\xi}t}}(t\geq0)$,
where $\Gamma^\xi$ is the filtering width~\cite{genes2008robust,gut2020stationary}.
Here, $F_t^{\xi}$ is responsible for filtering the TMS temporal modes at early time, while $F_b^\xi$ takes the BS temporal modes at later time.
From the standard input-output relation ${\hat a_\xi^{out}}(t) = \sqrt {\kappa^\xi}  \hat a_\xi (t)- {\hat a_\xi^{in}}(t)$, the selected output mode of the $\xi$-th independent levitated optomechanical system is given as $\hat a_\xi^{\mathrm{f}} (t)= \int {{F_{t(b)}^\xi}\left( {t - s} \right)} {\hat a_\xi^{out}}\left( s \right)\mathrm{d}s$~\cite{gut2020stationary}. Based on the above discussion,
it is possible to induce the motional entanglement between two distant levitated optomechanical systems by the Bell-like detection, in which the filtered outputs are interfaced by the beam splitter, then the conjugate homodyne detections are applied as shown in Fig.~\ref{Fig:1}~(a)~\cite{spedalieri2013covariance,SM,eghbali2015generating}.
In this way, the steady unconditional entanglement between two distant levitated nanoparticles can be generated ~\cite{spedalieri2013covariance}.

$\emph{Results}.-$
In order to investigate the entanglement between the output temporal output mode and the motional mode, and the entanglement between two distant motional modes, 
we adopt the logarithmic negativity $En$ to evaluate the steady entanglement~\cite{vitali2007optomechanical,pirandola2006macroscopic,abdi2012entanglement}
 \begin{equation}
 {E_n} \equiv \max \left[ {0, - \ln 2{\eta ^ - }} \right],
 \label{Eq:5}
 \end{equation}
 where ${\eta ^ - } \equiv \left( {{1 \mathord{\left/
 {\vphantom {1 {\sqrt 2 }}} \right.
 \kern-\nulldelimiterspace} {\sqrt 2 }}} \right){\left[ {\sum {\left( V \right) - \sqrt {\sum {{{\left( V \right)}^2}}  - 4\det V} } } \right]^{{1 \mathord{\left/
 {\vphantom {1 2}} \right.
 \kern-\nulldelimiterspace} 2}}}$  is the smallest partially-transposed symplectic eigenvalue of CM $V \equiv \left\{ {{A_1},{A_3};A_3^T,{A_2}} \right\}$ and $\sum {\left( V \right)}  \equiv \det A_1 + \det A_2 - 2\det A_3$.
$A_1$, $A_2$, and $A_3$ are the $2\times2$ block matrixes of V.
As an example, we calculate the entanglement for torsional motions in the following, while the entanglement for CoM motion has the similar result. 

\begin{figure}
  \centering
  \includegraphics[height=6cm]{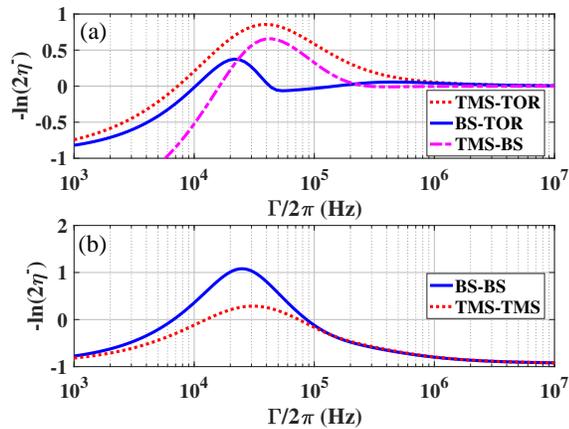}
  \caption{(Color online) $-ln(2\eta^{-})$ as a function of the filtering width $\Gamma$ in a single levitated system (a) and the two remote levitated system (b).
  In picture (a), the legend TMS (BS)-TOR denotes the output entanglement between the filtered TMS (BS) temporal mode and the torsional mode, while
  the legend TMS-BS represents the entanglement between the filtered TMS and BS temporal modes.
  From picture (b),
  The legend BS-BS (TMS-TMS) denotes that the entanglement between torsional modes induced by measuring the filtered output BS (TMS) temporal modes after the beam splitter. 
  Parameters are listed in this picture: the optical tweezers power $P_t=0.01$ W, the pressure of residual gas $P=10^{-4}$ Pa, temperature of residual gas $T_a=300$ K, the temperature for torsional mode $T=300$ K, the accommodation efficient $\gamma_{ac}=0.9$, the quantum efficiency $\eta=1$.
  Other parameters are the same with Fig.~\ref{Fig:2}.}
 \label{Fig:3}
\end{figure}

Based on the theory of CM  $V^{out}$ (cf., Sec.~IV in ~\cite{SM}) and Eq.~\eqref{Eq:5},
we investigate the steady entanglement between the output temporal modes and torsional mode in a single levitated optomechanical system. 
Previously, most of studies focused on the coupling regime where the rotating-wave approximations was valid. 
Therefore, the BS temporal mode cannot entangle with the torsional mode (cf., Sec.~VII in~\cite{SM,genes2008robust}).
In the ultrastrong coupling regime, both the rotating-wave terms and counter-rotating wave terms are involved to generate the entanglement.
As shown in Fig.~\ref{Fig:3} (a),
the steady entanglement can be generated between the BS/TMS temporal mode and torsional mode, and between the BS and TMS temporal modes.

The generated steady entanglement also depends on the filtering width $\Gamma$, which is in verse proportional to the interaction time between the torsional modes and the cavity modes. 
$\Gamma$ cannot be too large, or the incoherent spectral components would be collected and the interaction time is not enough to generate the entanglement~\cite{gut2020stationary}.  
The optimum $\Gamma$ is related to the effective decay rate of the torsional mode, which would approximately fit the width of Stokes and anti-Stokes peaks in output spectral~\cite{genes2008robust}.

Furthermore, we study the remote entanglement between two distant levitated nano-ellipsoids.
Without loss of generality,
we assume that the two levitated optomechanical systems are identical and the same filtering operations are performed to extract the TMS or BS temporal modes (parameters $s^{\xi}=s$). 
The steady entanglement between two distant torsional modes can be generated by Bell-like detection, and evaluated by CM $V_F$ (cf., Sec.~VI in~\cite{SM}) and Eq.~(\ref{Eq:5}).
%
As depicted in Fig.~\ref{Fig:3} (b), the Bell-like detection of BS temporal modes induced entanglement between two distant torsional modes is larger than the entanglement induced by measuring the TMS temporal modes. This seems contradict to the results shown in Fig.~\ref{Fig:3} (a), where the entanglement of TMS-TOR is lager than that that of BS-TOR~\cite{pirandola2006macroscopic,abdi2012entanglement,borkje2011proposal}.
However, as we known,  most photons are scattered into the BS temporal mode for the red sideband ($\Delta=\omega_\phi$).  In output cavity modes, the density of the state for the BS temporal mode is much greater than that of the TMS temporal mode~\cite{aspelmeyer2014cavity}.
Therefore, the optimum method to induce the steady entanglement between the two distant nano-ellipsoids is to measure the filtered BS temporal modes. 
\begin{figure}[htbp]
  \includegraphics[height=6cm]{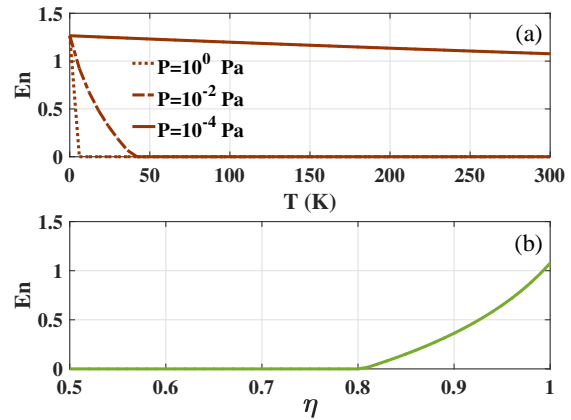}
  \caption{(Color online) (a) The steady entanglement of two distant torsional modes $E_n$ as a function of the environment temperature $T$ with different pressure of the residual gas $P$. (b) The steady entanglement of two distant torsional modes $E_n$ as a function of the quantum efficiency $\eta$.
  The BS temporal modes of each independent levitated optomechanical system are filtered with $\Gamma=1.5\times10^5$ rad$\cdot$Hz. Other parameters are the same with Fig.~\ref{Fig:3}.}
 \label{Fig:4}
\end{figure}

In our simulation, the decay rate of torsional motion mainly depends on the drag of the residual gas~\cite{millen2020optomechanics,hoang2016torsional,halbritter1974torque}, while the photon recoiling on torsional mode is neglected~\cite{windey2019cavity}.
According to the theoretical estimation~\cite{halbritter1974torque,hoang2016torsional}, the quality factor of the torsional exceeds $10^9$ if the pressure of the residual gas reaches $P=10^{-4}$ Pa (cf., Sec.~VII in \cite{SM}).
It enables the system to have a very long coherence time~\cite{hoang2016torsional,SM}.
Therefore, as shown in Fig.~\ref{Fig:4}~(a), the steady entanglement of two distant torsional modes is robust to the temperature of the thermal bath.
Even under the room temperature, the relatively high quantum entanglement between two torsional modes can still achieve.

The above results are based on an ideal Bell-like detection with quantum efficiency $\eta=1$, and without signal loss.  Theoretically, the signal loss can be included in a beam splitter with the non-ideal transmissivity ($0<\eta<1$)~\cite{leonhardt1997measuring}. As shown in Fig.~\ref{Fig:4}~(b),  the steady entanglement between the two distant torsional modes decreases to zero when the quantum efficiency $\eta$ is around $0.8$.
In practical systems, the quantum efficiency is mainly affected by signal losses in both the optical fiber and the detectors. In order to estimate the practical quantum efficiency $\eta$, we assume that the average attenuation of the optical fiber is 0.14 dB/km and detection efficiency of the superconducting detector is $98\%$ of wavelength $1550$ nm~\cite{reddy2020superconducting,tamura2018first}. 
The maximum distance between the two entangled nano-ellipsoids could be $12$ km.

$\emph{Conclusion}.-$
In summary, we have proposed a scheme to achieve the  steady motional entanglement between two distant optically levitated nanoparticles in a unconditional way.
The ultra-strong coupling between
the motion of the nanoparticle and 
the cavity mode is feasible under the coherent scattering mechanism.
Therefore, the rotating-wave approximation is no longer valid here. 
The optical tweezers are in red sideband of the cavity modes, and the strong and robust entanglement between the nanoparticle and the output cavity mode can be generated.
Furthermore, by the Bell-like detection, two distant nanoparticles can be entangled unconditionally through entanglement swapping.
The generated remote steady entanglement is a valuable resource for quantum network~\cite{Fiaschi2021}, quantum precise measurement \cite{2014NatPh..10..582K}, etc. 
Besides, in the ultra-strong optomechanical coupling regime, a lot of new phenomena and applications become possible, such as the single photon blockade \cite{PhysRevLett.107.063601,PhysRevA.101.063802}, quantum simulations \cite{2019NatRP,zhang2021quantum}, etc.

\begin{acknowledgements}
This research is supported by National Natural Science Foundation of China under Grant No. 61771278 and Beijing Institute of Technology Research Fund Program for Young Scholars.
\end{acknowledgements}

\bibliography{main.bib}

\end{document}


\title{Supplemental Material for \\ Steady motional entanglement between two distant levitated nanoparticles}

\author{Guoyao Li}
	\affiliation{Center for Quantum Technology Research and Key Laboratory of Advanced Optoelectronic Quantum Architecture and Measurements (MOE), School of Physics, Beijing Institute of Technology, Beijing 100081, China}

\author{Zhang-qi Yin}\email{zqyin@bit.edu.cn}
	\affiliation{Center for Quantum Technology Research and Key Laboratory of Advanced Optoelectronic Quantum Architecture and Measurements (MOE), School of Physics, Beijing Institute of Technology, Beijing 100081, China}
	\affiliation{Beijing Academy of Quantum Information Sciences, Beijing 100193, China}
	
\date{\today}
\maketitle
\renewcommand{\theequation}{S\arabic{equation}}
\tableofcontents
\thispagestyle{empty}
\newpage
\section{Polarization approximation in nano-ellipsoid}
\setcounter{page}{1}
An uniform isotropic non-dispersive nano-ellipsoid with semiaxes $a>b>c$ is assumed much smaller than the wavelength of the incoming optical field $\hat E(r,t)$. 
In this limit, the dipole approximation is applied. Furthermore, 
the polarization of the light field is assumed to be paralleled to one of the principal axes of the nano-ellipsoid, i.e., along axis $j$ $(j=a,b,c)$. 
The polarizability of nano-ellipsoid is given as \cite{bohren2008absorption}
 \begin{equation}
{\alpha _j} = 4\pi abc{\varepsilon _0}\frac{{{\varepsilon _r} - {\varepsilon _0}}}{{3{\varepsilon _0} + 3{L_j}\left( {{\varepsilon _r} - {\varepsilon _0}} \right)}}
\label{A1}
 \end{equation}
where
 \begin{equation}
{L_j} = \frac{{abc}}{2}\int_0^\infty  {\frac{{ds}}{{\left( {s + {j^2}} \right){{\left( {s + {a^2}} \right)}^{{1 \mathord{\left/
 {\vphantom {1 2}} \right.
 \kern-\nulldelimiterspace} 2}}}{{\left( {s + {b^2}} \right)}^{{1 \mathord{\left/
 {\vphantom {1 2}} \right.
 \kern-\nulldelimiterspace} 2}}}{{\left( {s + {c^2}} \right)}^{{1 \mathord{\left/
 {\vphantom {1 2}} \right.
 \kern-\nulldelimiterspace} 2}}}}}}.
 \label{A2}
 \end{equation}
Here $\varepsilon _r$ and $\varepsilon _0$ are the permittivities of the nano-ellipsoid and the vacuum.
With $a>b=c$, the $L_j$ can be written as 
\begin{equation}
\begin{array}{l}
{L_a} = \frac{{1 - {e^2}}}{{{e^2}}}\left( { - 1 + \frac{1}{{2e}}\ln \frac{{1 + e}}{{1 - e}}} \right),\\
\\
{L_a} + {L_b} + {L_c} = 1,\\
\\
{L_b} = {L_c}
\end{array}
 \end{equation}
 where $e = \sqrt {1 - {{{b^2}} \mathord{\left/
 {\vphantom {{{b^2}} {{a^2}}}} \right.
 \kern-\nulldelimiterspace} {{a^2}}}} $ denotes eccentricity.
Without loss of efforts, the above equations can be extended to other shapes like prolate spheroid, oblate spheroid, and etc~\cite{bohren2008absorption}.

\begin{figure}
  \centering
  \includegraphics[height=8cm]{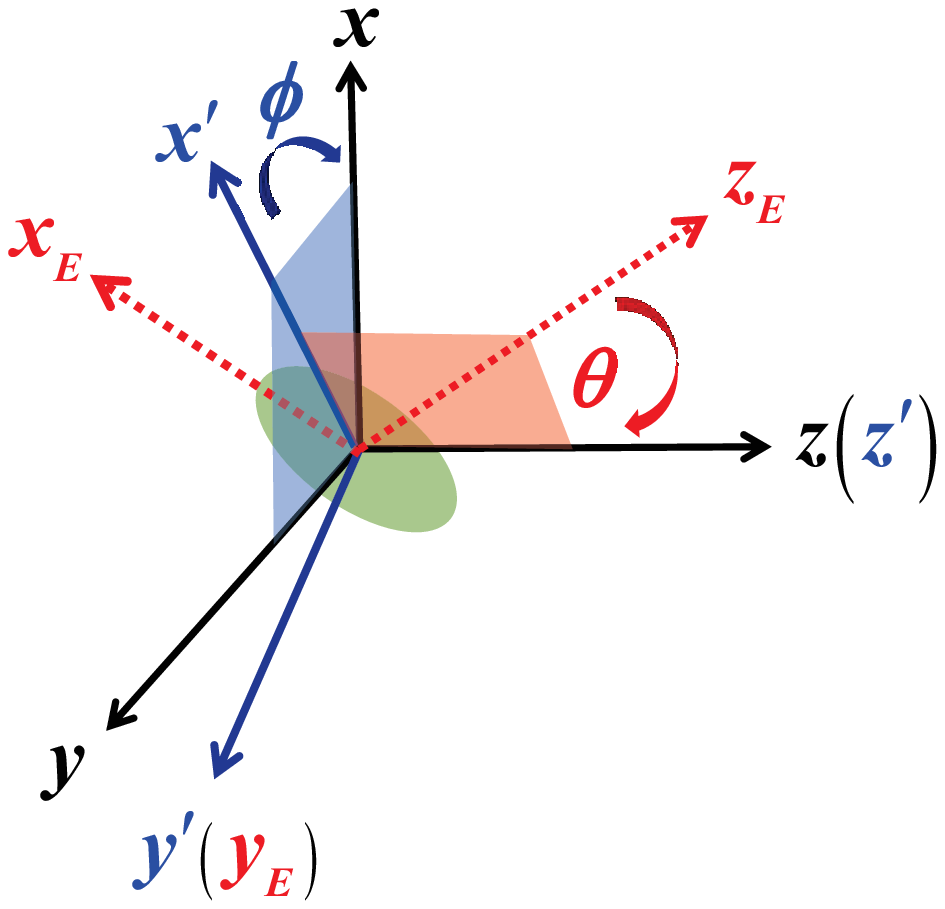}
  \caption{(Color online) The scheme for the orientation of the ellipsoid. By rotating the axis $y_E$ and $z'$ with the Euler angles $\theta$ and $\phi$, the Cartesian coordinate system ${\{x_E,y_E,z_E\}}$ of the ellipsoid maps to a fix experimental Cartesian coordinates system ${\{x,y,z\}}$.}
  \label{F1}
\end{figure}

In an optical field, the induced dipole moment of the ellipsoid is the sum of dipole moments in three principal axes.
The corresponding polarizability can be written as a tensor
\begin{equation}
\hat{\alpha } = \left| {\begin{array}{*{20}{c}}
{{\alpha _a} }&0&0\\
0&{{\alpha _b} }&0\\
0&0&{{\alpha _c} }
\end{array}} \right|.
\end{equation}
 Then a transform matrix is used to cast the polarizability from principal axes $\{x_E,y_E,z_E\}$ to the experimental coordinate system $\{x,y,z\}$ by the Euler angles $\{\theta,\phi\}$ as shown in Fig.~(\ref{F1}).
 The polarizability tensor of the rotated nano-ellipsoid can be rewritten in the form of matrix product~\cite{rudolph2021theory}
 \begin{equation}
\hat{ \alpha} ' = {\hat{R}^{ - 1}}\hat{\alpha} \hat{R}
 \end{equation}
 where
 \begin{equation}
 \hat{R} = \left| {\begin{array}{*{20}{c}}
{\cos \theta }&0&{ - \sin \theta }\\
0&1&0\\
{\sin \theta }&0&{\cos \theta }
\end{array}} \right|\left| {\begin{array}{*{20}{c}}
{\cos \phi }&{\sin \phi }&0\\
{ - \sin \phi }&{\cos \phi }&0\\
0&0&1
\end{array}} \right|.
  \end{equation}
The induced dipole moment of the nano-ellipsoid in experimental coordinate system $\{x,y,z\}$
can be written as
\begin{equation}
\hat{p} = \hat{\alpha}'(\theta,\phi)\hat{E}(r,t).
\end{equation}

\section{Electric field in optical cavity}\label{B}

In a single levitated optomechanical system, the nano-ellipsoid is trapped by optical tweezers with frequency $\omega_0$. 
The levitated nano-ellipsoid scatters the photon of the optical tweezers into the cavity mode with frequency $\omega_c$. 
The electric field acting on the nano-ellipsoid can be read by $\hat{E}\left( {r,t} \right) = {\varepsilon _{tw}}\left( {r,t} \right) + {\hat{E}_c}\left( r \right)$ where
 \begin{equation}
{\varepsilon _{tw}}\left( {r,t} \right) = \frac{1}{2}\left[ {{E_t}\left( r \right){e^{i{\omega _0}t}} + c.c.} \right]
 \end{equation}
describes the mean value of the optical tweezers field and
 \begin{equation}
{\hat{E}_c} = \sqrt {\frac{{\hbar {\omega _c}}}{{2{\varepsilon _0}{V_c}}}} \left[ {f\left( r \right)\hat{a} + H.c.} \right]
 \end{equation}
is the electric field of the cavity mode.
$\hat a^{\dag}$ and $\hat{a}$ are the bosonic creation and annihilation operators.
${V_c} = \pi L w_c^2/4$ is the cavity mode volume, where $L$ and $w_c$ are the cavity length and beam waist respectively.
Limited by the boundary of the plane-parallel mirrors, the mode function $f(r)$ in $y$ direction can be simplified as $f\left( y \right) = \cos \left( {{k_c}y + \varphi } \right)$ where
$k_c$ ($\varphi$) is the wave number (phase) of the cavity field.

Generally, the optical tweezers field can be approximately described as a Gaussian form
${E_t}(r)= {E_0}\frac{{{w_0}}}{{w\left( z \right)}}{e^{ - \frac{{{x^2} + {y^2}}}{{{w^2}\left( z \right)}}}}{e^{i{\omega _0}t}}{e^{ik_tz}}{e^{i{\psi _t}\left( r \right)}}$ where ${E_0} = \sqrt {\frac{{4{P_t}}}{{\pi {\varepsilon _0}cw_0^2}}} $ and  $w_0$ are the amplitude and beam waist of the tweezers in the focus.
$w\left( z \right) = {w_0}\sqrt {1 + {{{z^2}} \mathord{\left/
 {\vphantom {{{z^2}} {z_R^2}}} \right.
 \kern-\nulldelimiterspace} {z_R^2}}} $ is the tweezers waist along the propagation $z$. ${z_R} = {{\pi w_0^2} \mathord{\left/
 {\vphantom {{\pi w_0^2} \lambda_t }} \right.
 \kern-\nulldelimiterspace} \lambda_t}$ is the Rayleigh range in Gaussian beam. ${\psi _t}\left( r \right)$ is the well-know Gouy phase of the tweezers. $P_t$, $c$, $\lambda_t$, and $k_t$ are the power, speed, wavelength, and wave number of the optical tweezers respectively.
This field yields a Gaussian-shaped optical potential well which provides a harmonic trap in the vicinity of the maximum field amplitude.

\section{Torsional and Center-of-Mass modes of the nano-ellipsoid}
\begin{figure}
  \centering
  \includegraphics[height=8cm]{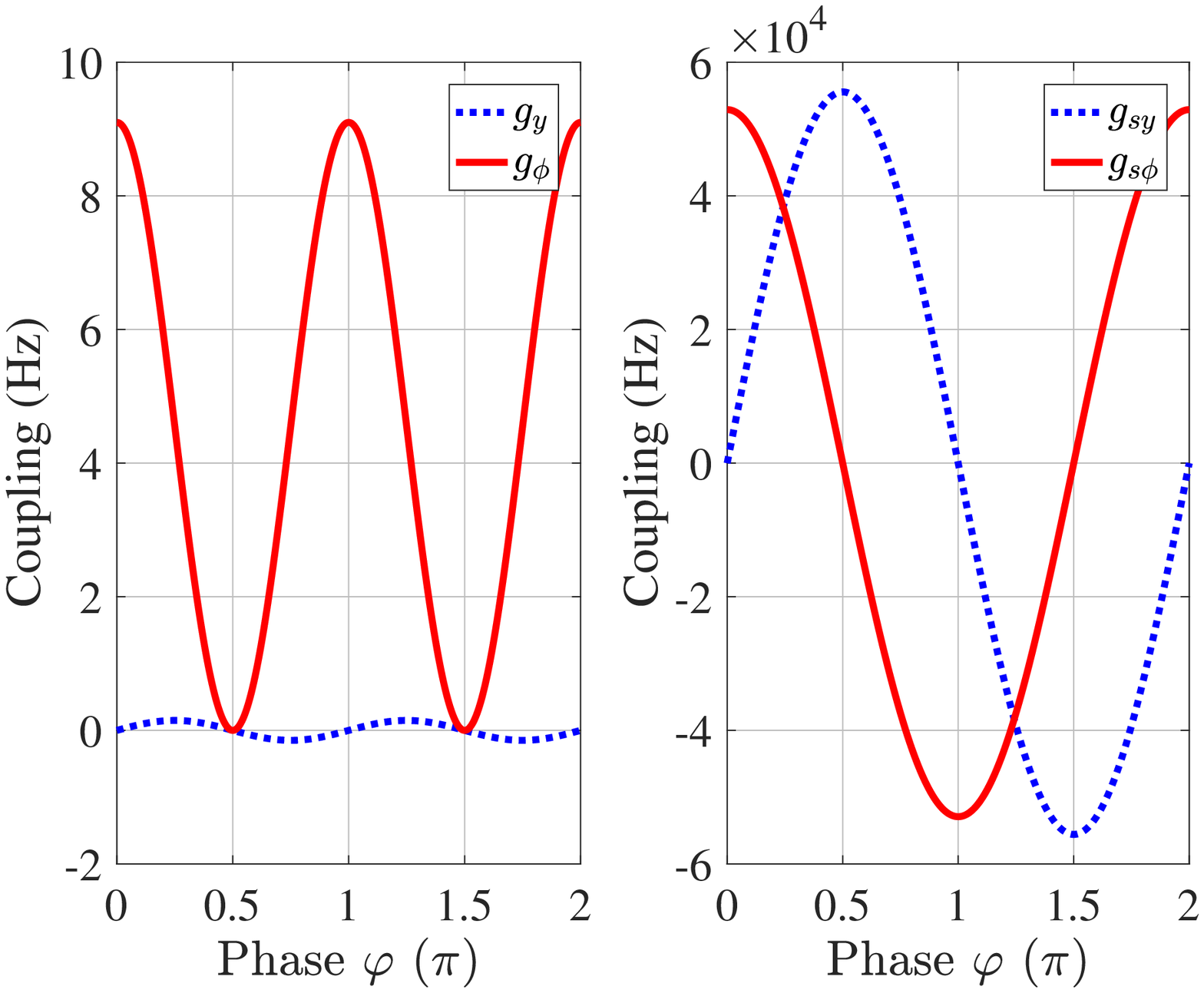}
  \caption{(Color online) The coupling strength $(g_y,g_\phi,g_{sy},g_{s\phi})$ as a function of the cavity phase $\varphi$. The parameters are the same with Fig. 2 in maintext.}
  \label{S2}
\end{figure}
In a single levitated optomechanical system, the interaction Hamiltonian between the levitated nano-ellipsoid and electromagnetic field can be written as \cite{gonzalez2019theory}
 \begin{equation}
{\hat{H}_I} = -\frac{1}{2}\hat{\alpha}'(\theta,\phi) {\hat{E}^2(r,t)}.
 \end{equation}
It can be divided into three parts:
 \begin{equation}
 {\hat{H}_{t - t}} = -\frac{1}{2}\hat{\alpha} '\varepsilon _{tw}^2\left( {r,t} \right),
 \end{equation}
  \begin{equation}
 {\hat{H}_{c - c}} = -\frac{1}{2}\hat{\alpha} '\hat{E}_{c}^2(r),
 \end{equation}
  \begin{equation}
 {\hat{H}_{t - c}} = -\hat{\alpha} '\varepsilon _{tw}(r,t)\hat{E}_c(r),
 \end{equation}
where $\hat{H}_{t-t}$ denotes the interaction between the optical tweezers, $\hat{H}_{c-c}$ describes the interaction between the cavity fields, and $\hat{H}_{t-c}$ represents the coherent scattering interaction between the optical tweezers and cavity field.

With assumption of $\theta,\phi=0$, 
only CoM motion in experimental coordinate system $\{x,y,z\}$ is under considered.
One can expand $\hat{H}_{t-t}$ to the second order of $r$.
The frequency of CoM motion in $y$ direction is given by 
\begin{equation}
{\omega _y} = \sqrt {{{E_0^2{\alpha _a}} \mathord{\left/
 {\vphantom {{E_0^2{\alpha _a}} {mw_0^2}}} \right.
 \kern-\nulldelimiterspace} {mw_0^2}}},
 \end{equation} where $m$ is the mass of the nano-ellipsoid.
 The corresponding coupling can be attained by expending the Hamiltonians $\hat{H}_{c-c}$ and $\hat{H}_{t-c}$ to the first order of $r$, listed in order by 
optomechanical coupling
\begin{equation}
{g_y} = {{{\alpha _a}{\omega _c}{k_c}{y_0}\sin 2\varphi } \mathord{\left/
 {\vphantom {{{\alpha _a}{\omega _c}{k_c}{y_0}\sin 2\varphi } {2{\varepsilon _0}{V_c}}}} \right.
 \kern-\nulldelimiterspace} {2{\varepsilon _0}{V_c}}}
 \end{equation}
 and coherent scattering coupling 
 \begin{equation}
 {g_{sy}} = {\alpha _a}{E_0}{k_c}{y_0}\sin \varphi \sqrt {{{{\omega _c}} \mathord{\left/
 {\vphantom {{{\omega _c}} {2\hbar {\varepsilon _0}{V_c}}}} \right.
 \kern-\nulldelimiterspace} {2\hbar {\varepsilon _0}{V_c}}}} 
 \end{equation}
 respectively. Here $y_0$ is the zero-point fluctuation of CoM mode. The cavity mode frequency is already modified as ${\omega _c} = {\omega _c}\left( {1 - \frac{{{\alpha _1}}}{{2{\varepsilon _0}{V_c}}}{{\cos }^2}\varphi } \right)$.

On the other hand, we can also make a assumption that the nano-ellipsoid is fixed at the origin of the experimental coordinates, i.e., $x,y,z=0$.
One can expend $\hat{H}_{t-t}$ to the second order of $\theta$ $(\phi)$ as $\phi  \to 0$ $\left( {\theta  \to 0} \right)$, where the angular potential is harmonic to the Euler angle $\theta$ $(\phi)$.
The corresponding frequency of the torsional motion is given by
\begin{equation}
{\omega _{\theta(\phi) }} = {E_0}\sqrt {\frac{{{\alpha _a} - {\alpha _b}}}{{2I}}}
\end{equation}
where $I = {{m\left( {{a^2} + {b^2}} \right)} \mathord{\left/
 {\vphantom {{m\left( {{a^2} + {b^2}} \right)} 5}} \right.
 \kern-\nulldelimiterspace} 5}$ is inertia of the nano-ellipsoid.
 Similarly, the optomechanical coupling and coherent scattering coupling between cavity mode and torsional motion can be calculated by expending $H_{c-c}$ and $H_{t-c}$ to the first order of $\theta$ ($\phi$) as $\phi  = 0,\theta  \to {\pi  \mathord{\left/
 {\vphantom {\pi  4}} \right.
 \kern-\nulldelimiterspace} 4}$ $\left( {\theta  = 0,\phi  \to {\pi  \mathord{\left/
 {\vphantom {\pi  4}} \right.
 \kern-\nulldelimiterspace} 4}} \right)$, listed in order by
\begin{equation}
 {g_{\theta (\phi)}} = \frac{{\left( {{\alpha _a} - {\alpha _b}} \right){\omega _c}{\phi _0}{{\cos }^2}\left( \varphi  \right)}}{{2{\varepsilon _0}{V_c}}}
\end{equation}
and
\begin{equation}
 {g_{s\theta(s\phi) }} = \left( {{\alpha _a} - {\alpha _b}} \right){E_0}{\phi _0}\cos \left( \varphi  \right)\sqrt {\frac{{{\omega _c}}}{{8\hbar {\varepsilon _0}{V_c}}}}.
\end{equation}
${\phi _0} = \sqrt {{\hbar  \mathord{\left/
 {\vphantom {\hbar  {2I{\omega _\phi }}}} \right.
 \kern-\nulldelimiterspace} {2I{\omega _\phi }}}}$
is the zero-point fluctuation of torsional mode.
The cavity mode frequency is amended by ${\omega _c} = {\omega _c}\left( {1 - \frac{{{\alpha _1} + {\alpha _2}}}{{4{\varepsilon _0}{V_c}}}{{\cos }^2}\varphi } \right)$.

The above results are consistent with that of the perturbation theory~\cite{hoang2016torsional}. As shown in Fig.~\ref{S2},
by adjusting the position of the nano-ellipsoid to the antinode (node) of the cavity field $\varphi=n\pi/2$ ($n=0,1,2,...$), the torsional mode and CoM can be decoupled from each other.
Besides, the $g_{\phi}$ ($g_y$) is smaller enough than $g_{s\phi}$ ($g_{sy}$), we mainly consider the coherent scattering coupling $g_{s\phi}$ ($g_{sy}$) in this letter.

\section{Dynamics of the output field}
Without loss of generality, we take the torsional motion $\phi$ as a example in the following. 
The Hamiltonian of a single levitated optomechanical system can be written in the interaction picture:
\begin{equation}
\hat{H} = \hbar \Delta {\hat{a}^\dag }\hat{a} + \hbar {\omega _\phi }{\hat{b}^\dag_\phi }\hat{b }_\phi- {g_{s\phi }}\left( {{\hat{a}^\dag } + \hat{a}} \right)\left( {{\hat{b}^\dag_\phi } + \hat{b}_\phi} \right)
\label{S17}
\end{equation}
 where $\Delta=\omega_c-\omega_0$. $\hat{a},\hat{a}^\dag$ $(\hat{b}_\phi,\hat{b}^\dag_\phi)$ are the annihilation and creation operators of the cavity mode (torsional mode) satisfying $\left[ {\hat{a},{\hat{a}^\dag }} \right] = 1$ $(\left[ {\hat{b}_\phi,{\hat{b}^\dag_\phi }} \right] = 1)$.
 The first two terms on right side represent the free Hamiltonian of the cavity mode and torsional mode, while the last term denotes the coherent scattering interaction between the cavity mode and torsional mode.

Derived from Eq.~\ref{S17}, the linearized Langevin equations are given as follow:
\begin{equation}
\dot {\hat{a}} =  - i\Delta \hat{a} - \frac{\kappa }{2}\hat{a} + ig_{s\phi}\left( {\hat{b}_\phi + {\hat{b}^\dag_\phi }} \right) + \sqrt k {\hat{a}^{in}},
\label{S18}
\end{equation}
\begin{equation}
\dot {\hat{b}}_\phi =  - i{\omega _\phi }\hat{b}_\phi - \frac{\gamma }{2}\hat{b}_\phi + ig_{s\phi}\left( {{\hat{a}^\dag } + \hat{a}} \right) + \sqrt {\gamma_\phi}  {\hat{b}^{in}}_\phi,
\label{S19}
\end{equation}
where $\kappa$ and $\gamma_\phi$ are the decay rates of the cavity and torsional modes.
The operators are already replaced by $\delta \hat{o} \to \hat{o}$ $(\hat{o}=\hat{a},\hat{b}_\phi)$, in which $\delta \hat{o}$ represents the fluctuation of each operator.
The zero-mean Gaussian noises of cavity mode and torsional mode are $\hat{a}^{in}$ and $\hat{b}^{in}_\phi$,  satisfying correlation functions~\cite{vitali2007optomechanical}
\begin{equation}
\left\langle {{\hat{a}^{in}}\left( t \right){\hat{a}^{in\dag }}\left( {t'} \right)} \right\rangle  = \delta \left( {t - t'} \right)
\end{equation}
and
\begin{equation}
\left\langle {{\hat{b}^{in}_\phi}\left( t \right){\hat{b}^{in\dag }_\phi}\left( {t'} \right)} \right\rangle  = \left( {\bar n_\phi + 1} \right)\delta \left( {t - t'} \right)
\end{equation}
where
$\bar n_\phi = {\left( {\exp \left\{ {{{\hbar {\omega _\phi }} \mathord{\left/
 {\vphantom {{\hbar {\omega _\phi }} {{k_B}T}}} \right.
 \kern-\nulldelimiterspace} {{k_B}T}}} \right\} - 1} \right)^{ - 1}}$
 is the mean thermal excitation number for torsional frequency $\omega_{\phi}$ at temperature $T$. $k_B$ is the Boltzmann constant.

Next, we define the dimensionless quadratures $\hat{Q} = {\hat{b}^\dag_\phi } + \hat{b}_\phi,\hat{P }= i\left( {{\hat{b}^\dag_\phi } - \hat{b}_\phi} \right),\hat{X} = {\hat{a}^\dag } + \hat{a},\hat{Y}= i\left( {{\hat{a}^\dag } - \hat{a}} \right)$, and the corresponding noise quadratures ${\hat{Q}^{in}} = {\hat{b}^{in\dag }_\phi} + {\hat{b}^{in}_\phi},{\hat{P}^{in}} = i\left( {{\hat{b}^{in\dag }_\phi} - {\hat{b}^{in}_\phi}} \right),{\hat{X}^{in}} = {\hat{a}^{in\dag }} + {\hat{a}^{in}},{\hat{Y}^{in}} = i\left( {{\hat{a}^{in\dag }} - {\hat{a}^{in}}} \right)$. 
The Eqs.~\eqref{S18} and \eqref{S19} can be rewritten as a compact matrix form
\begin{equation}
\hat{\dot {u}}( t ) = A\hat{u}( t) + \hat{n}( t ),
\end{equation}
where ${\hat{u}^T} (t)= \left \{ {\hat{Q}(t),\hat{P}(t),\hat{X}(t),\hat{Y}(t)} \right \}$, ${\hat{n}^T(t)} = \left \{ {\sqrt {\gamma_\phi} {\hat{Q}^{in}(t)},\sqrt {\gamma_\phi } {\hat{P}^{in}(t)},\sqrt \kappa  {\hat{X}^{in}(t)},\sqrt \kappa  {\hat{Y}^{in}(t)}} \right \}$, and the drift matrix
\begin{equation}
A = \left| {\begin{array}{*{20}{c}}
{ - \frac{\gamma_\phi }{2}}&{{\omega _\phi }}&0&0\\
{ - {\omega _\phi }}&{ - \frac{\gamma_\phi }{2}}&{2g_{s\phi}}&0\\
0&0&{ - \frac{\kappa }{2}}&\Delta \\
{2g_{s\phi}}&0&{ - \Delta }&{ - \frac{\kappa }{2}}
\end{array}} \right|.
\end{equation}

Then, it is convenient to characterize the fluctuation of two-mode Gaussian state by the $ 4\times4 $ covariance matrix (CM) $V$, whose components are given by ${V_{i,j}} = {{\left\langle {{u_i}\left( \infty  \right){u_j}\left( \infty  \right) + {u_j}\left( \infty  \right){u_i}\left( \infty  \right)} \right\rangle } \mathord{\left/
 {\vphantom {{\left\langle {{u_i}\left( \infty  \right){u_j}\left( \infty  \right) + {u_j}\left( \infty  \right){u_i}\left( \infty  \right)} \right\rangle } 2}} \right.
 \kern-\nulldelimiterspace} 2}$ $(i,j=1,2,3,4)$.
 In the steady state, the corresponding CM $V$ of intra cavity can be numerically calculated by
 $AV+VA^{T}=D$ where $D = diag\left \{ {\frac{\gamma }{2}\left( {2\bar n_\phi + 1} \right),\frac{\gamma }{2}\left( {2\bar n _\phi+ 1} \right),\frac{\kappa }{2},\frac{\kappa }{2}} \right \}$ denotes the correlations of different noise~\cite{vitali2007optomechanical}.
Here, we concern about the output CM. The output temporal modes with different frequency should be selected as we interest.
Such operation can be done by introducing a filter
which enables to decompose the continuous traveling lights into different temporal modes and code it in a time sequence. The filtering operation can be depicted mathematically by \cite{gut2020stationary}
\begin{equation}
{F_t} = \sqrt {2\Gamma } {e^{\Gamma t}}{e^{ - i{\omega _\phi}t}}(t\leq0),
\end{equation}
\begin{equation}
{F_b} = \sqrt {2\Gamma } {e^{ - \Gamma t}}{e^{i{\omega _\phi}t}}(t\geq0),
\end{equation}
where $\Gamma$ is the filtering width.
$F_t$ extracts the TMS temporal modes at early time while $F_b$ gets the BS temporal modes at later time.
Therefor, one can selected the output modes by the filtering operation $\hat a^{f}(t) = \int {{F_{t(b)}}\left( {t - s} \right)} {\hat a^{out}}\left( s \right)ds$. 
Combining with the standard input-output relation ${\hat a^{out}}(t) = \sqrt \kappa  \hat a(t) - {\hat a^{in}(t)}$, the output CM $V^{out}$ is given by \cite{gut2020stationary,genes2008robust}
\begin{equation}
{V^{out}} = \int {T\left( \omega  \right)} S\left( \omega  \right)DS{\left( { - \omega } \right)^T}T{\left( { - \omega } \right)^T}d\omega
\label{Eq:26}
\end{equation}
where $S(\omega)=CM(\omega)+P$ with $M\left( \omega  \right) = {\left( {i\omega I + A} \right)^{ - 1}}$,
\begin{equation}
C = \left| {\begin{array}{*{20}{c}}
1&0&0&0\\
0&1&0&0\\
0&0&{\sqrt \kappa  }&0\\
0&0&0&{\sqrt \kappa  }\\
0&0&{\sqrt \kappa  }&0\\
0&0&0&{\sqrt \kappa  }
\end{array}} \right|,
\end{equation}
and
\begin{equation}
P = \left| {\begin{array}{*{20}{c}}
0&0&0&0\\
0&0&0&0\\
0&0&{{1 \mathord{\left/
 {\vphantom {1 {\sqrt \kappa  }}} \right.
 \kern-\nulldelimiterspace} {\sqrt \kappa  }}}&0\\
0&0&0&{{1 \mathord{\left/
 {\vphantom {1 {\sqrt \kappa  }}} \right.
 \kern-\nulldelimiterspace} {\sqrt \kappa  }}}\\
0&0&{{1 \mathord{\left/
 {\vphantom {1 {\sqrt \kappa  }}} \right.
 \kern-\nulldelimiterspace} {\sqrt \kappa  }}}&0\\
0&0&0&{{1 \mathord{\left/
 {\vphantom {1 {\sqrt \kappa  }}} \right.
 \kern-\nulldelimiterspace} {\sqrt \kappa  }}}
\end{array}} \right|.
\end{equation}
 $I$ is the identity matrix.
$T(\omega)$ is the Fourier transform of
\begin{equation}
T\left( t \right) = \left| {\begin{array}{*{20}{c}}
{\delta \left( t \right)}&0&0&0&0&0\\
0&{\delta \left( t \right)}&0&0&0&0\\
0&0&{{R_t}}&{ - {I_t}}&0&0\\
0&0&{{I_t}}&{{R_t}}&0&0\\
0&0&0&0&{{R_b}}&{ - {I_b}}\\
0&0&0&0&{ - {I_b}}&{{R_b}}
\end{array}} \right|
\end{equation}
where $R_t$ and $I_t$ ($R_b$ and $I_b$) are the real and imaginary parts of $F_t$ ($F_b$).

\section{Output entanglement in a single levitated optomechanical system}
\begin{figure}
  \centering
  \includegraphics[height=8cm]{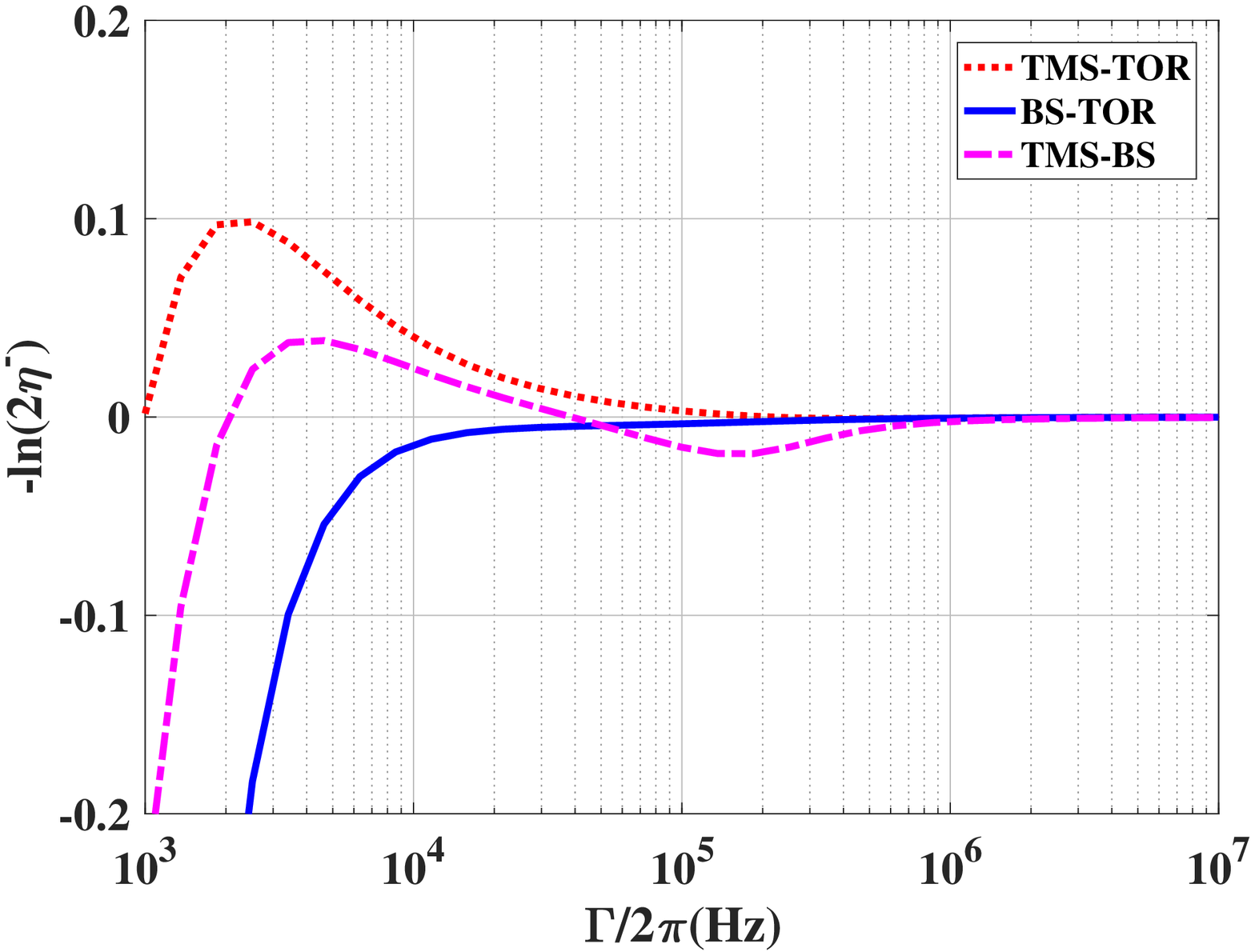}

  \caption{(Color online) The $-ln(2\eta^{-})$ as a function of the filtering width $\Gamma$ in a single levitated optomechanical system. The legend TMS(BS)-TOR denotes the output entanglement between the filtered TMS(BS) temporal mode and torsional mode. Another legend TMS-BS represents the output entanglement between the TMS and BS temporal modes. The coherent scattering coupling ($g_{s\phi}=0.04\omega_\phi$) in this picture is one order of magnitude lower than the Fig.~3(a) ($g_{s\phi}=0.4\omega_\phi$) in main text, while other parameters are the same. Noted that the parameters in this letter fulfill the demand of steady trapping according to the Mie scattering theory~\cite{nieminen2007optical} and the Routh-Hurwitz criterion~\cite{vitali2007optomechanical}.}
  \label{S3}
\end{figure}
A continuous variable system is a canonical infinite dimensional quantum system with continuous eigenspectra, which can be characterized by $N$ bosonic modes.
The corresponding  Hilbert space is ${H^{ \otimes N}} =  \otimes _{k = 1}^N{H_k}$, where $H_k$ is spanned by the Fock basis $\left\{ {\left| n \right\rangle } \right\}_{n = 0}^\infty $ of eigenstates with the number operator ${\hat n_k} = \hat a_k^\dag {\hat a_k}$. $\hat a_k$ and $\hat a_k^\dag$ are the bosonic annihilation and creation operators, yielding the commutation relations $\left[ {{{\hat a}_i},\hat a_j^\dag } \right] = {\delta _{i,j}}$.
Those operators can be arranged to the quadrature field operators ${\hat q_k} = {\hat a_k} + \hat a_k^\dag$ and ${\hat p_k} =  - i\left( {{{\hat a}_k} - \hat a_k^\dag } \right)$ via the Cartesian decomposition.
It associates to the position and momentum operators of the quantum harmonic oscillator.
The vector $\hat x = {\left( {{{\hat q}_1},{{\hat p}_1},...,{{\hat q}_N},{{\hat p}_N}} \right)^T}$ must obey
$\left[ {{{\hat x}_i},{{\hat x}_i}} \right] = 2i{\Omega _{ij}}$ where
$\Omega  = \mathop  \oplus \limits_{k = 1}^n \left( {\begin{array}{*{20}{c}}
0&1\\
{ - 1}&0
\end{array}} \right)$ is the symplectic form.
A quantum state $\hat{\rho}$ of $N$-mode bosonic system is Gaussian if its Winger representation is Gaussian~\cite{weedbrook2012gaussian}, 
which can be fully described by the displacement vector $\vec{x}=Tr(\hat{x}\hat{\rho})$ and the $2N\times2N$ CM 
$V_{ij}=\langle\{\Delta\hat{x}_{i},\Delta\hat{x}_{j}\}\rangle/2$.
$\{,\}$ is the anticommutator and $\Delta\hat{x}_i=\hat{x}_i-\langle\hat{x}_i\rangle$.
The quantum entanglement of two-mode Gaussian state can be evaluated by the well-know logarithmic negativity~\cite{vitali2007optomechanical}
 \begin{equation}
 {En} \equiv \max \left[ {0, - \ln 2{\eta ^ - }} \right],
 \label{Eq:43}
 \end{equation}
 where ${\eta ^ - } \equiv \left( {{1 \mathord{\left/
 {\vphantom {1 {\sqrt 2 }}} \right.
 \kern-\nulldelimiterspace} {\sqrt 2 }}} \right){\left[ {\sum {\left( V \right) - \sqrt {\sum {{{\left( V \right)}^2}}  - 4\det V} } } \right]^{{1 \mathord{\left/
 {\vphantom {1 2}} \right.
 \kern-\nulldelimiterspace} 2}}}$  is the smallest partially-transposed symplectic eigenvalue of CM $V \equiv \left\{ {{A_1},{A_3};A_3^T,{A_2}} \right\}$ and $\sum {\left( V \right)}  \equiv \det A_1 + \det A_2 - 2\det A_3$. $A_1$, $A_2$, and $A_3$ are the $2\times2$ block matrixes of V.

The levitated nano-ellipsoid and cavity field are initial in thermal state and vacuum state respectively.
Driving by the optical tweezers, the levitated optomechanical system can be characterized by the multi-mode Gaussian state if the system being linear and stable~\cite{aspelmeyer2014cavity}.
The corresponding steady output entanglement in a levitated optomechanical system can be evaluated by Eqs.~\ref{Eq:26} and \ref{Eq:43}.
As shown in Fig.~\ref{S3}, $-ln(2\eta^{-})$  as a function of the filtering width $\Gamma$ in weak coupling regime ($g_{s\phi}=0.04\omega_\phi$). 
With the state swapping by the beam-splitter interaction, the output entanglement of TMS-BS are induced. Noted that the BS interaction will not contribute to the entanglement of BS-TOR in weak coupling~\cite{genes2008robust}.

\section{Bell-like detection}
The Bell-like detection is used to measure the CM of $n+2$ bosonic modes \cite{spedalieri2013covariance}. 
In this method, the last two modes A and B are interfaced by a beam splitter with arbitrary transmissivity $T_r$.
Then the inputs will be transformed to new output modes CM. 
Measured by two conjugate homodyne detecting respectively. 
the CM of the rest $n$ bosonic modes can be obtained.

Based on the output CM of two levitated optomechanical system, the CM of the whole system is given by
\begin{equation}
{V_T} = \left| {\begin{array}{*{20}{c}}
{E}&C\\
{{C^T}}&O
\end{array}} \right|
\end{equation}
where
\begin{equation}
E = \left| {\begin{array}{*{20}{c}}
{{V_A}\left( {1:2;1:2} \right)}&{J}\\
{J}&{{V_B}\left( {1:2;1:2} \right)}
\end{array}} \right|
\end{equation}
is the reduced CM of torsional modes.
$V_A$ and $V_B$ are the CM of levitated optomechanical systems A and B, which can be calculated by Eq.~\ref{Eq:26}.
Similarly,
\begin{equation}
O = \left| {\begin{array}{*{20}{c}}
{{V_A}\left( {3:4;3:4} \right)}&{Z}\\
{Z}&{{V_B}\left( {3:4;3:4} \right)}
\end{array}} \right|
\end{equation}
is the reduced CM of the cavity modes for each levitated optomechanical system. 
Here, $4\times4$ matrix $J$ $(Z)$ represents the cross correlation of torsional modes (cavity modes), the elements of which are 0 if two levitated optomechanical systems are independent.
Matrix
\begin{equation}
C = \left| {{V_A}\left( {1:2;3:4} \right),{V_B}\left( {1:2;3:4} \right)} \right|
\end{equation}
is a rectangular matrix showing the correlation between the torsional mode and cavity mode.
For convenience, we define the matrixes as
\begin{equation}
\begin{array}{l}
{V_A}(3:4;3:4) = \left| {\begin{array}{*{20}{c}}
{{a_1}}&{{a_3}}\\
{{a_3}}&{{a_2}}
\end{array}} \right|,\\
\\
{V_B}(3:4;3:4) = \left| {\begin{array}{*{20}{c}}
{{b_1}}&{{b_3}}\\
{{b_3}}&{{b_2}}
\end{array}} \right|,\\
\\
Z = \left| {\begin{array}{*{20}{c}}
{{z_1}}&{{z_3}}\\
{{z_4}}&{{z_2}}
\end{array}} \right|.
\end{array}
\end{equation}

Generally, the signals are not ideal since the signal modes inevitably couple to the environment. 
The loss in optical fiber and detectors can be included to the beam splitter with non-ideal transmissivity $0<\eta<1$. 
The quantum efficiency of $l$-th detector can be written as $\eta_l=\eta^0_l 10 \time 10 ^{-\alpha_0 d/10}$ where $\eta^0_l$ is the intrinsic detector efficiency~\cite{PhysRevA.88.022336}.
$\alpha_0$ ($d$) is the loss coefficient (length) of the channel.
In terms of the non-ideal Bell-like detection,
the retaining CM of torsional modes is given by \cite{spedalieri2013covariance}
\begin{equation}
{V_F} = E - \frac{1}{{\det \Upsilon }}\sum\limits_{i,j = 1}^2 {{C_i}} {K_{ij}}C_j^T,
\label{Eq:35}
\end{equation}
where
\begin{equation}
\Upsilon  = \left| {\begin{array}{*{20}{c}}
{{r_1}}&{{r_3}}\\
{{r_3}}&{{r_2}}
\end{array}} \right|
\end{equation}
with
\begin{equation}
\begin{array}{l}
{r_1} = \left( {1 - T_r} \right){a_1} + T_r{b_1} - 2\sqrt {T_r\left( {1 - T_r} \right)} {z_1} + \frac{{1 - {\eta _1}}}{{{\eta _1}}},\\
\\
{r_2} = \left( {1 - T_r} \right){b_2} + T_r{a_2} - 2\sqrt {T_r\left( {1 - T_r} \right)} {z_2} + \frac{{1 - {\eta _2}}}{{{\eta _2}}},\\
\\
{r_3} = \sqrt {T_r\left( {1 - T_r} \right)} \left( {{b_3} - {a_3}} \right) - \left( {1 - T_r} \right){z_3} + T_r{z_4}.
\end{array}
\end{equation}
The elements of matrix $K$ are listed by
\begin{equation}
\begin{array}{l}
{K_{11}} = \left| {\begin{array}{*{20}{c}}
{\left( {1 - T_r} \right){r_2}}&{\sqrt {\left( {1 - T_r} \right)T_r} {r_3}}\\
{\sqrt {\left( {1 - T_r} \right)T_r} {r_3}}&{T_r{r_1}}
\end{array}} \right|,\\
\\
{K_{22}} = \left| {\begin{array}{*{20}{c}}
{T_r{r_2}}&{ - \sqrt {\left( {1 - T_r} \right)T_r} {r_3}}\\
{ - \sqrt {\left( {1 - T_r} \right)T_r} {r_3}}&{\left( {1 - T_r} \right){r_1}}
\end{array}} \right|,\\
\\
{K_{12}} = K_{21}^T = \left| {\begin{array}{*{20}{c}}
{ - \sqrt {\left( {1 - T_r} \right)T_r} {r_2}}&{\left( {1 - T_r} \right){r_3}}\\
{ - T_r{r_3}}&{ - \sqrt {\left( {1 - T_r} \right)T_r} {r_1}}
\end{array}} \right|.
\end{array}
\end{equation}

\section{Damping of nano-ellipsoid motion}

In the high vacuum, the damping of the torsional motion of the levitated nano-ellipsoid is mainly dominated by the collision of surrounding rare gas. 
The damping of the torsional motion can be derived from the change of the angular momentum, given by~\cite{halbritter1974torque}
\begin{equation}
\begin{aligned}
\gamma_\phi  = \frac{{5{\rho _a}\bar v{a^3}\sqrt {1 - {e^2}} }}{{8\rho \left( {{a^2} + {b^2}} \right)}}\left[ {{\gamma _{ac}}\left( {{f_1} + \left( {1 - {e^2}} \right){f_2}} \right) + 3\left( {1 - {\gamma _{ac}}\frac{{6 - \pi }}{8}} \right){e^4}{f_3}} \right]
\end{aligned}
\end{equation}
where
\begin{equation}
\begin{array}{l}
{f_1} = \frac{3}{{8{e^2}}}\left[ {\frac{1}{e}\arcsin \left( e \right) - \left( {1 - 2{e^2}} \right)\sqrt {1 - {e^2}} } \right],\\
\\
{f_2} = \frac{3}{{16{e^2}}}\left[ {\left( {1 + 2{e^2}} \right)\sqrt {1 - {e^2}}  - \frac{1}{e}\arcsin \left( e \right)\left( {1 - 4{e^2}} \right)} \right],\\
\\
{f_3} = \frac{1}{{4{e^4}}}\left[ {\left( {3 - 2{e^2}} \right)\sqrt {1 - {e^2}}  + \frac{1}{e}\arcsin \left( e \right)\left( {4{e^2} - 3} \right)} \right].
\end{array}
\end{equation}
Similarly, the damping for the CoM motion is
\begin{equation}
{\gamma _y} = \frac{{3{\rho _a}\bar va}}{{8\rho {b^2}}}\left[ {\frac{{{\gamma _{ac}}}}{2}\left( {1 - {e^2} + \sqrt {1 - {e^2}} \arcsin \left( e \right)/e} \right) + \left( {1 - {\gamma _{ac}}\left( {6 - \pi } \right)/8} \right)\left( {\left( {1 - {e^2}} \right)/{e^2} + \left( {2{e^2} - 1} \right)\sqrt {1 - {e^2}} \arcsin \left( e \right)/{e^3}} \right)} \right].
\end{equation}
${\rho _a} = {{{m_a}{P}} \mathord{\left/
 {\vphantom {{{m_a}{p_a}} {{k_B}}}} \right.
 \kern-\nulldelimiterspace} {{k_B}}}{T_a}$
is the mass density, where $m_a$, $P$, and $T_a$ are atom mass, pressure, and temperature of the residual gas.
$\bar v = \sqrt {{{8{k_B}{T_a}} \mathord{\left/
 {\vphantom {{8{k_B}{T_a}} {\pi {m_a}}}} \right.
 \kern-\nulldelimiterspace} {\pi {m_a}}}} $ denotes the mean thermal velocity and $\rho$ is the density of the nano-ellipsoid.
$ {\gamma _{ac}}$ represents the accommodation efficient which relates to the diffuse and specular reflection of ellipsoidal surface. With parameters of Fig.~2 in main text, the damping rate for torsional (CoM) motion is $\gamma_\phi=9.1\times10^{-5}$~Hz ($\gamma_y=4.1\times10^{-4}$~Hz) if the pressure of the residual gas reaches $P=10^{-4}$~Pa. The corresponding quality factor of torsional (CoM) motion achieves $Q_\phi=\omega_\phi/\gamma_\phi=1.4\times10^{9}$ ($Q_y=\omega_y/\gamma_y=3.3\times10^{8}$).

\bibliography{supp.bib}